\documentstyle[aps,preprint,epsf]{revtex}
\begin{document}
\tighten
\preprint{DESY 94-165,
{\small E}N{\large S}{\Large L}{\large A}P{\small P}-A-487/94,
TUW-94-18
}
\draft
\title{Effects of weak self-interactions in a relativistic plasma\\
       on cosmological perturbations}
\author{Herbert Nachbagauer\thanks{e-mail: herby@lapphp1.in2p3.fr}
}
\address{{Laboratoire de Physique Th\'eorique}
ENSLAPP
\thanks{URA 14-36 du CNRS, associ\'ee \`a l'E.N.S. de Lyon, et au L.A.P.P.
(IN2P3-CNRS) d'Annecy-le-Vieux.}
\\
B.P. 110, F-74941 Annecy-le-Vieux Cedex, France}
\author{Anton K. Rebhan\thanks{On leave of absence from
        Institut f\"ur Theoretische Physik der
         Technischen Universit\"at Wien;
         e-mail: rebhana@x4u2.desy.de}
         }
\address{DESY, Gruppe Theorie,\\
        Notkestra\ss e 85, D-22603 Hamburg, Germany}
\author{Dominik J. Schwarz
        \thanks{e-mail: dschwarz@ecxph.tuwien.ac.at}
        }
\address{Institut f\"ur Theoretische Physik,
         Technische Universit\"at Wien,\\
         Wiedner Hauptstra\ss e 8-10/136,
         A-1040 Wien, Austria}

\date{November 30, 1994}

\maketitle

\begin{abstract}
The exact solutions for linear cosmological perturbations
which have been obtained for collisionless relativistic matter
within thermal field theory are extended to a self-interacting case.
The two-loop contributions of scalar $\lambda\phi^4$
theory to the thermal graviton self-energy are evaluated, which give
the $O(\lambda)$ corrections in the perturbation equations.
The changes are found to be perturbative on scales
comparable to or larger than the Hubble horizon, but 
the determination of
the large-time damping behavior of subhorizon
perturbations requires a resummation of thermally induced masses.
\end{abstract}

\pacs{PACS numbers: 98.80.-k, 11.10.Wx, 52.60.+h}

\narrowtext

The theory of linear perturbations of otherwise
homogeneous and isotropic cosmological models \cite{Lifshitz}
plays a central r\^ole in the problem of large-scale structure
formation in the early universe.
The conventional approach to their study in
cosmological models involving more complicated
forms of matter than a perfect fluid \cite{Lifshitz} is based on
the coupled Einstein-Boltzmann equations
\cite{Peebles70,Ehlers,Stewart72,Peebles73,Bond83},
which are usually solved
numerically. In Ref.\ \cite{Kraemmer}, a novel framework
has been developed which employs
perturbative thermal field theory \cite{Landsman}
to derive self-consistent and automatically gauge-invariant
perturbation equations. In this formalism,
the connection of the
perturbed metric and the perturbed energy-momentum tensor is provided
by the thermal graviton self-energy. At one-loop order, the leading
high-temperature contributions to the latter have been obtained in
Ref.\ \cite{Rebhan91} (see also Ref.\ \cite{BFT})
in momentum space, and conformal invariance
allows one to transform these results directly
to a curved space with vanishing Weyl tensor.
The one-loop high-temperature contributions
describe relativistic collisionless matter interacting with the
metric perturbations, and the resulting perturbation equations have
been shown \cite{AKRDS} to be equivalent to a gauge-invariant
reformulation of the Einstein-Vlasov equations\cite{Kasai}. Moreover,
analytic solutions for scalar, vector, and tensor perturbations
have been found for a spatially flat
Friedmann-Robertson-Walker (or, Einstein-de Sitter) cosmological model
in the general case of a two-component system containing also a
perfect fluid part \cite{Kraemmer,AKRDS,Rebhan92b}, as well
as for higher-dimensional models \cite{Schwarz}.

In this Letter, we investigate the effects of weak self-interactions
on the exact solutions that were found for relativistic
collisionless matter. For simplicity we consider scalar particles
with $\lambda\phi^4$ interaction Lagrangian. The leading
self-interaction
effects are described by two-loop corrections to the thermal graviton
self-energy in the high-temperature limit.
The corrections are such that the ensuing perturbation equations
can still be solved exactly in terms of rapidly converging
power series.

The background space-time of the Einstein-de Sitter cosmological
model, which we shall consider, is given through the line element 
\begin{equation}
ds^2 = S^2(\tau) \left( -d\tau^2 + \delta_{ij}dx^i dx^j \right)
\label{ds2}\ ,
\end{equation}
with $\tau$ being the conformal time which measures the size
of the horizon in comoving coordinates, $R_H=\tau$.
A thermal distribution of
relativistic (i.e., effectively massless) scalar particles with
$\lambda\phi^4$ self-interactions gives rise to
a background energy-momentum tensor
\begin{equation}
\tilde{T}^{\mu}{}_{\nu} = u^{\mu}u_{\nu} (\tilde{E} + \tilde{P}) +
\tilde{P} \delta^{\mu}_{\nu}\ , \qquad u_{\mu} = S\delta_\mu^0 \ ,
\label{eit}
\end{equation}
with $\tilde{E} = 3\tilde{P}$. Including order $\lambda$,
$\tilde{T}^{\mu}{}_{\nu}$ is determined by the
diagrams of Fig.~1, which at finite temperature $T$ yield 
\begin{equation}
\tilde{E}(S=1)=\left(\frac{\pi^2}{30}-\frac\lambda{16}\right)T^4
  + O(\lambda^{3/2}) \ .
\label{eit00}
\end{equation}
The scale dependence of the mean energy density is given by
$\tilde{E}(S) = \tilde{E}(S=1) S^{-4}$,
with $S\propto\tau$.


Self-consistent
perturbations of the metric $g_{\mu\nu}=\tilde g_{\mu\nu}
+\delta g_{\mu\nu}$ are found by varying the Einstein equations,
\begin{equation}\label{pEeq}
\delta G^{\mu\nu}\equiv \delta(R^{\mu\nu}-\frac12 g^{\mu\nu}R)=
-8\pi G \delta T^{\mu\nu}
\end{equation}
with
\begin{equation}
\delta T_{\mu\nu}(x)=\int d^4y
{\delta T_{\mu\nu}(x)\over \delta g^{\alpha\beta}(y)}
\delta g^{\alpha\beta}(y).
\end{equation}
Hence, $\delta T_{\mu\nu}$ is determined by the graviton self-energy
\begin{equation}
\Pi_{\mu\nu\alpha\beta}\equiv
{\delta^2\Gamma\over\delta g^{\mu\nu}
\delta g^{\alpha\beta}}=\frac12{\delta(\sqrt{-g}T_{\mu\nu})\over
\delta g^{\alpha\beta}},
\label{pi}
\end{equation}
where $\Gamma$ contains all contributions to the effective action
besides the classical Einstein-Hilbert action.

$\Gamma$ is a nonlocal functional of the metric. With a flat background
metric, the graviton self-energy $\Pi_{\mu\nu\alpha\beta}$ is
most easily evaluated through momentum-space Feynman rules.
In the high-temperature limit, $\Gamma$ is conformally invariant
\cite{Kraemmer}
which makes it possible to directly transform the flat-space
results to conformally flat cosmological models such as the
Einstein-de Sitter one of Eq.~(\ref{ds2}) according to
\begin{eqnarray}
\label{ftpi}
\left. \Pi^{\mu\nu\rho\sigma}(x,x^{\prime})
\right|_{g=S^2\eta} &&  \nonumber \\
= S^{-2}(\tau)
\int  {d^4 k \over (2\pi)^4}  & &
e^{\imath k (x - x^{\prime})} \left.
\bar{\Pi}^{\mu\nu\rho\sigma}(k)\right|_{\eta} S^{-2}(\tau^{\prime}) \ .
\end{eqnarray}

The one-loop contribution to $\bar{\Pi}^{\mu\nu\rho\sigma}$
has been evaluated in Ref.\ \cite{Rebhan91}
at temperature $T\gg k_0,k$, which is relevant for the case of
cosmological perturbations [the momentum scale is set by the
inverse Hubble radius $\sim (T/m_{\rm
\rm Planck})T$]. 
It describes collisionless
matter, interacting solely with the metric perturbations.
Since collisionless matter can sustain anisotropic pressure,
this medium admits a rich spectrum of possible cosmological solutions.
Those have been studied in great detail in Ref.\ \cite{AKRDS}
by two of the present authors.

In order to investigate the effects of weak
self-interactions, one has to include higher loop
corrections.
With $\lambda\phi^4$ interactions,
the order $\lambda$ contributions are contained in the two-loop
diagrams shown in Fig.~2. Evaluating again the high-temperature
limit $T\gg k_0,k$
(full details will be presented elsewhere)
gives a contribution to $\bar{\Pi}^{\mu\nu\rho\sigma}(k)$
that again satisfies the Ward identity corresponding to
conformal invariance of $\Gamma$ which is necessary for
the simple transformation law of Eq.~(\ref{ftpi}).


Since we did not have to single out a particular gauge,
the perturbed Einstein equations (\ref{pEeq}) involve only gauge
invariant combinations of the perturbed metric components and
can be given entirely in terms of
Bardeen's gauge invariant potentials\cite{Bardeen}.
Expanding the latter in Fourier modes
(plane waves with respect to the conformally flat background
of Eq.~(\ref{ds2})) then leads
to equations depending only on the conformal time $\tau$, which
can be combined with the wavelength of the metric perturbations
$\ell \equiv 2\pi/k$
into the dimensionless variable
\begin{equation}
x \equiv k\tau = \frac{2\pi R_{\rm H}}\ell \ .
\end{equation}
$x/\pi=1$ defines the point in conformal time when half a wavelength
fits inside the growing Hubble horizon.

Because of the nonlocality of the effective action $\Gamma$, the
resulting equations are integro-differential equations rather
than ordinary differential equations, involving a convolution
integral of $\Pi^{\mu\nu\rho\sigma}(x,x^{\prime})$ with the
metric perturbations.

In the case of scalar metric perturbations, there are two
independent equations for the two gauge invariant
potentials $\Phi_H$ and $\Phi_A$ of Bardeen \cite{Bardeen}.
Using instead the combinations
\begin{equation}
\Phi = \frac12 \Phi_H, \qquad \Pi=-(\Phi_H+\Phi_A),
\end{equation}
which are 
related to the  gauge-invariant (cf. \cite{Bardeen})
density contrast  $\epsilon_{\rm m} = x^2 \Phi /3 $
and to the anisotropic pressure $\pi_{\rm anis} = x^2 \Pi$,  
one equation (the $00$-component of Eq.\ (\ref{pEeq}))
is of the form
\begin{eqnarray}
\FL
\label{scalar}
 (x^2-3)\Phi(x)+3x\Phi'(x)-6 \Pi(x)=    \nonumber \\
-12 \int dx' K(x-x') (\Phi+\Pi)'(x') \ ,
\end{eqnarray}
while the second is an ordinary differential equation by virtue of
the vanishing of the
trace of the energy-momentum tensor in our ultrarelativistic setting,
\begin{equation}
\label{trace}
\Phi^{\prime\prime} + {4\over x} \Phi^{\prime} +
{1\over 3}\Phi + {2\over x} \Pi^{\prime} - {2\over 3} \Pi=0 \ .
\end{equation}
All the remaining components of the perturbed Einstein equations
(\ref{pEeq}) turn out to be combinations of Eqs.\ (\ref{scalar}) and
(\ref{trace}).

The one-loop contribution to the kernel $K$
is determined by the discontinuity of the one-loop graviton
self-energy $\bar{\Pi}^{0000}(k_0,k)$
in Eq.~(\ref{ftpi}).
It reads\cite{Kraemmer} ($\omega = k_0 / k$)
\begin{eqnarray}\label{K1}
K_{1}(x-x')=\frac{i}{2\pi}
\int\limits_{-\infty+i\varepsilon}^{\infty+i\varepsilon}
d\omega e^{-i\omega(x-x')}\frac12\ln\frac{\omega+1}{\omega-1}
\nonumber \\
= j_0(x-x')\theta(x-x'),
\end{eqnarray}
with $j_0(x)\equiv \sin(x)/x$.
The step function gives an upper bound of $x$ in the integral in
Eq.~(\ref{scalar}); a lower bound at some initial $x_0$ can
be adopted by adding an inhomogeneous term
$\sum_{n=0}^\infty \gamma_n K^{(n)}(x-x_0)$ to the convolution
integral in Eq.~(\ref{scalar}), encoding the
infinite number of initial conditions one has to
impose by specifying the likewise infinite number of
moments of the initial particle distribution \cite{AKRDS}.

The two-loop contributions have more complicated structure, but
can still be evaluated exactly,
\begin{eqnarray}\label{K2}
K_{2}(x-x'&)&=\frac{5i\lambda}{32\pi^3}
\int\limits_{-\infty+i\varepsilon}^{\infty+i\varepsilon}
d\omega\,e^{-i\omega(x-x')} \nonumber \\
&& \quad \quad  \; \; \times \left[
\omega \ln^2\frac{\omega+1}{\omega-1}-
 \ln\frac{\omega+1}{\omega-1}-\frac{2\omega}{\omega^2-1}
\right] \nonumber \\
 & = & \frac{5\lambda}{8\pi^2} ( 2\kappa'-j_0-\cos )(x-x')\theta(x-x') \ ,
\end{eqnarray}
where we
introduced
\begin{eqnarray}
\kappa(x)&\equiv&
\frac1x\left[ \sin(x) {\rm Si}(2x)
+\cos(x) \{ {\rm Ci}(2x)-\gamma-\ln(2x) \}\right] \nonumber \\
&=&2\sum_{m=0}^\infty \frac{(-1)^m x^{2m+1}}{(2m+2)!}
\sum_{j=0}^m\frac1{2j+1} \ ,
\end{eqnarray}
with Si and Ci being the sine and cosine integral \cite{AS}.
 In Eq.\ (\ref{K2}), there are two contributions which
differ in form from the one-loop kernel: the first term in the
integrand, which arises
exclusively from the last diagram in Fig.~2, and the last term, which
is due solely to the third diagram.

With the explicit result for $K_2$, which has a well-behaved power
series representation, the same methods as in Ref.\ \cite{Kraemmer,AKRDS}
can be used to solve the coupled equations (\ref{scalar}) and
(\ref{trace}).
Since  $K_{2}(x)\propto x^4+O(x^6) $,
$K=K_{1}+K_{2}$ still satisfies
\begin{equation}
\label{K0}
K(0_+)=1,\qquad K(0_+)+3K''(0_+)=0 \ ,
\end{equation}
as did $K_{1}$. Thereby the conditions\cite{AKRDS}
on the initial data (summarized by the set of $\gamma_n$)
following from demanding regularity at $x=x_0=0$ are left unchanged.

In addition to regular solutions,
there exist also ones that are singular as $x\to0$.
In the collisionless case those behave as $\Phi,\Pi\sim
x^{-5/2}\cos[\sqrt{27/20}\ln(x)]$, thus
showing oscillations even on superhorizon
scales\cite{AKRDS,Zakharov79,Vishniac82}. This behavior is modified
by the weak self-interactions, to wit,
\begin{equation}
\label{ccrit}
\Phi,\Pi\sim
x^{-5/2}\cos[\sqrt{\frac{27}{20}-\frac{2\lambda}{\pi^2}}\ln(x)] \ ,
\end{equation}
up to a constant shift of phase.

In Ref.\ \cite{Kraemmer}, it was pointed out that the coupled system
of Eqs.~(\ref{scalar}) and (\ref{trace}) can be solved exactly for
$x_0=0$, yielding power series representations with infinite radius
of convergence. This still holds true after adding the self-interaction
term $K_{2}$. For the particular initial condition determined
by $\gamma_{n}=0$, $n\ge1$, the first few terms in the
regular solutions are given by
\begin{eqnarray}
\Phi(x)&=&
\gamma_0\biggl(
\frac{28-10\lambda/\pi^2}{5} \\ \nonumber&&\! \qquad-
\frac{4644-3735\lambda/\pi^2 + 350(\lambda/\pi^2)^2}
{315(54-5\lambda / \pi^2)}\,
x^2  \pm \cdots
\biggr),\\ \nonumber
\Pi(x)&=&  \gamma_0\biggl(-
\frac{4-5\lambda/\pi^2}{5} \\ \nonumber&&\! \qquad+
\frac{2808 - 5445\lambda/\pi^2 + 700(\lambda/\pi^2)^2}
{630(54-5\lambda / \pi^2)}\,
x^2 \mp \cdots
\biggr).
\end{eqnarray}
With $\lambda = 1$, which corresponds to a prefactor
$\approx 0.06$ in the last line of Eq.~(\ref{K2}),
the density contrast $\epsilon_{\rm m}=x^2\Phi/3$ is
plotted in Fig.~3 by the dashed line and compared with the
collisionless case (full line).
On superhorizon scales, $x\ll1$, the solutions are dominated by
the constant modes in $\Phi$ and $\Pi$. The amount of
anisotropic pressure associated with a given density contrast
is determined by $\Pi/\Phi$, which is seen to be reduced 
with increasing $\lambda$, in agreement with intuitive expectation,
since the collision-dominated 
case of a perfect fluid has $\Pi\equiv0$.


For large $x\gg1$, the asymptotic behavior
of the density contrast
$\epsilon_{\rm m}(x)$ is proportional to the one of $K(x)$. In the
collisionless case $K(x)=K_{1}(x)
\sim \sin(x)/x$, whereas in the other extreme
case of a perfect fluid $\epsilon_{\rm m}\sim \sin(x/\sqrt3)$.
With weak self-interactions, the asymptotic behavior is indeed
modified towards weaker damping (see Fig.\ 3 and 4),
which ceases completely for sufficiently large $x$ where
$K(x)\sim \lambda\cos(x)$. However, there the correction $K_{2}$
is overtaking the lowest order term $K_{1}$, which signals
a possible breakdown of perturbation theory.

Inspecting the analytic structure that gives rise to the different
asymptotic behavior of
$K_{2}$, one finds that the discontinuity in the integrand of
Eq.~(\ref{K2}) is singular on the light-cone $\omega 
=\pm1$.
There is a logarithmically singular term which is responsible
for $\kappa'(x)\sim \sin(x)\ln(x)/x$, but there are
even poles at $\omega=\pm1$, which contribute the term
involving the $\cos(x)$. However,
the scalar particles, whose rest masses are negligible
at sufficiently high temperature, acquire thermal masses
$m=\sqrt{\lambda}T$. A resummation of these those will remove the
singularities at $\omega=\pm1$, which are associated with
the masslessness of the scalar particles.

The logarithmically singular integrand of Eq.\ (\ref{K2}) will
be regularised at $\omega=\pm1$, but since the logarithmic singularity
is integrable anyway, the resummed result will not change dramatically;
the poles at $\omega=\pm1$, however, 
disappear completely after a resummation of the thermal masses.

Adding a thermal mass term to the Lagrangian and subtracting it again
as a higher-order counter-term has the effect of replacing all the
propagators in Fig.~2 by massive ones and subtracting out the third
and fourth diagram, which were precisely the ones contributing the pole
term to Eq.~(\ref{K2}). This resummation of mass insertions replaces
the leading-order result $K_1(x)= j_0(x)\theta(x)$ by
\begin{eqnarray}\label{K1r}
K_1^{\rm res.}(x)=- \theta(x)&& \int_{-1}^1 d\omega e^{-i\omega x}\\ \nonumber
\times
\int_{m/\sqrt{1-\omega^2}}^\infty && dp\,p^4
{d\over dp}\left({  1\over \exp (p/T) -1 } \right)
\left/ \left( {8\pi^4 T^4\over15} \right)  \right.
\end{eqnarray}
up to terms whose amplitude is down by factors of $\lambda$.
With a non-zero $m=\sqrt{\lambda}T$,
the integrand is now seen to vanish at $\omega=\pm1$, whereas
before it was a constant; in fact, it vanishes
together with all its derivatives, but rapidly recovers the bare one-loop
value away from the light-cone. This is a negligible
effect for small $x$,
but the large $x$ behavior is dominated by the (non-analytic) integration
region
$|\omega| \approx 1$.

Eq.\ (\ref{K1r}) can be evaluated by a Mellin transform\cite{Dav}
which yields
\begin{eqnarray}\label{K1rexp}
&&K_1^{\rm res.}(x)=\theta(x)
{\sqrt\pi\over2}{15\over 8 \pi^4} \times\\ \nonumber
&&\sum_{k=0}^\infty \lambda^{k/2}
{(-1)^{1+k}(k-4)\zeta(k-3)\over (2\pi)^{k-4} \Gamma(1+k/2)}
\left(2\over x\right)^{{1-k\over 2}} J_{{1-k \over 2}}(x) ,
\end{eqnarray}
where in the term with $k=4$ one has to substitute $(k-4)\zeta(k-3)\to1$.
For small $x$,
\begin{equation}\label{K1rexp1}
K_1^{\rm res.}(x)
=\theta(x) \left( j_0(x)-{5\lambda\over 8\pi^2}\cos(x) \right)
+O(\lambda^{3/2})
\end{equation}
is a good approximation;
for $x\gg1$, on the other hand, the complete function
$K_1^{\rm res.}(x)$ turns out to
decay even slightly faster than $j_0(x)$, oscillating with a reduced
phase velocity
\begin{equation}
v=1-{5\lambda\over 8\pi^2}+O(\lambda^{3/2}).
\end{equation}

A first approximation to a fully resummed calculation is therefore
to change
$$
K_{1}(x)\propto j_0(x) \to j_0((1-\frac{5\lambda}{8\pi^2})x),
$$
which reduces
$$
K_{2}(x)\to \frac{5\lambda}{4\pi^2} ( \kappa'-j_0 )(x)\theta(x)
$$
up to higher orders in $\lambda$.
However, the latter of the relations in Eq.~(\ref{K0}) is
now violated at the higher order $O(\lambda^2)$, which drastically
restricts the solvability of the perturbation equations,
so further modifications are needed. A natural solution
is to change the phase velocity in $K_{2}$ by the same amount
as in $K_{1}$ and to adjust the
prefactor in $K_{2}$ appropriately. This leads in a unique manner
to the remarkably simple result
\begin{equation}
\label{Kres}
K^{\rm res.}(x)=
\left( j_0+\frac{1-v^2}{v^2}(\kappa'-j_0)\right)(vx) \,\theta(x).
\end{equation}
Up to and including order $\lambda$,
this is the same as $K_{1}+K_{2}$. But now
the asymptotic behavior of $\epsilon_{\rm m}(x)$ for large $x$,
which is linked to $K(x)$, is $\sim \kappa'(vx) \sim \sin(vx)\ln(x)/x$.
Thus there
is still strong damping of subhorizon-scale oscillations, which
is only logarithmically weaker than in the collisionless case. Moreover,
the phase velocity $v$ is now smaller than 1, as one may expect
from the comparison with the strongly collisional case of a
perfect radiation fluid, where (undamped) acoustic waves oscillate
with $v=1/\sqrt3$.
In Fig.\ 4 the effects of the above resummation are shown for
$x/\pi \gtrsim 1$.


Similar issues arise also in the case of vector (rotational) as well
as tensor perturbations (primordial gravitational waves).
For small $x$, weak self-interactions
affect mainly the singular solutions, and the same shift of
frequency of the superhorizon oscillation occurs as in Eq.~(\ref{ccrit});
for large $x$, vector and tensor perturbations differ from the
scalar case in that their asymptotic behavior is not changed
as much. Subhorizon-scale vector perturbations still have
vorticity decaying $\sim\sin(x)/x$,
and resummation changes
only the phase velocity; tensor perturbations are stable against
the kind of resummation performed above, and have the same
asymptotic behavior as the collisionless ones,
which indeed is already the
same as in the perfect fluid case. All these findings as well as
a generalisation to more-component systems will be
presented in detail in a forthcoming publication. 

\acknowledgments

This work was supported by the Austrian ``Fonds zur F\"orderung der
wissenschaftlichen Forschung (FWF)'' under projects no.\ P9005-PHY
and P10063-PHY, and by the EEC Programme ``Human Capital
and Mobility'', 
contract CHRX-CT93-0357 (DG 12 COMA).

\begin{figure}
\bigskip\bigskip
\centerline{ \epsfxsize=5in \epsfbox[ 97.3728 343 502.697 430]{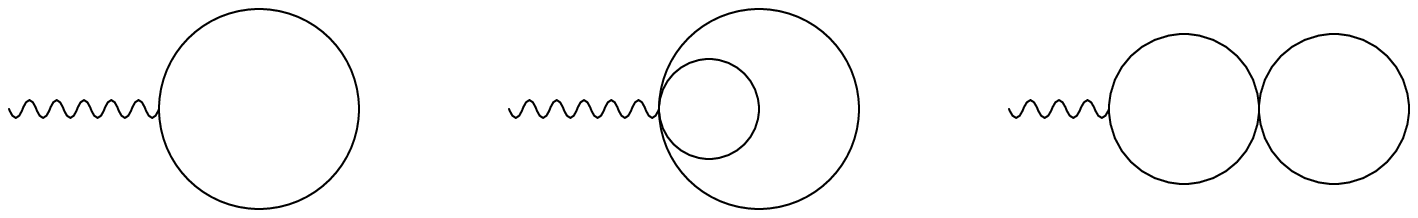} }
\caption{One- and two-loop diagrams for the energy-momentum tensor.
Wavy lines denote external gravitons and straight lines
scalar particles.}
\end{figure}

\begin{figure}
\centerline{ \epsfxsize=7in \epsfbox[33 320 564.6 505]{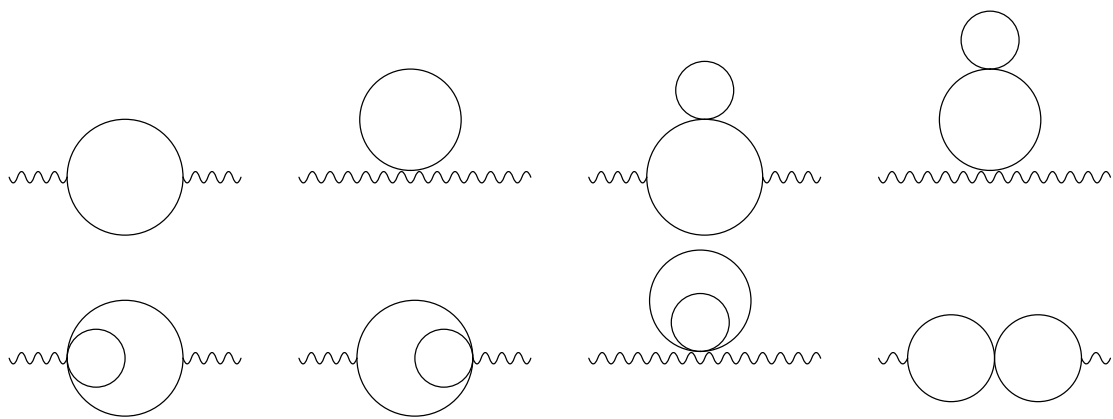} }
\caption{The one- and two-loop diagrams for the graviton
self-energy up to and including $O(\lambda)$. Only the last diagram
contributes to the
$\kappa^{\prime}$-term
in $K_{2}$.}
\end{figure}

\begin{figure}
\centerline{{ \epsfxsize=5in \epsfbox[72 155 540 640]{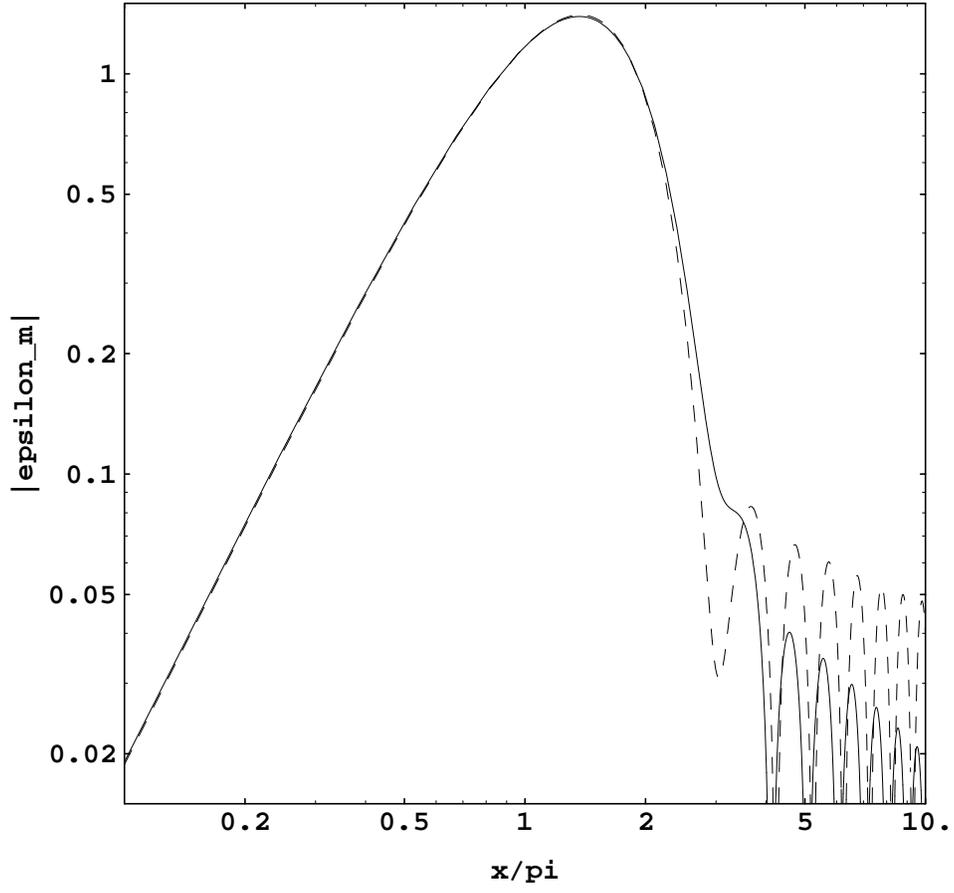} }}
\caption{The density contrast $|\epsilon_{\rm m}(x)|$
(in arbitrary normalization).
The full line shows the collisionless situation, the dashed
line includes self-interactions with $\lambda = 1$.}
\end{figure}

\begin{figure}
\centerline{{ \epsfxsize=6in \epsfbox[72 225 540 560]{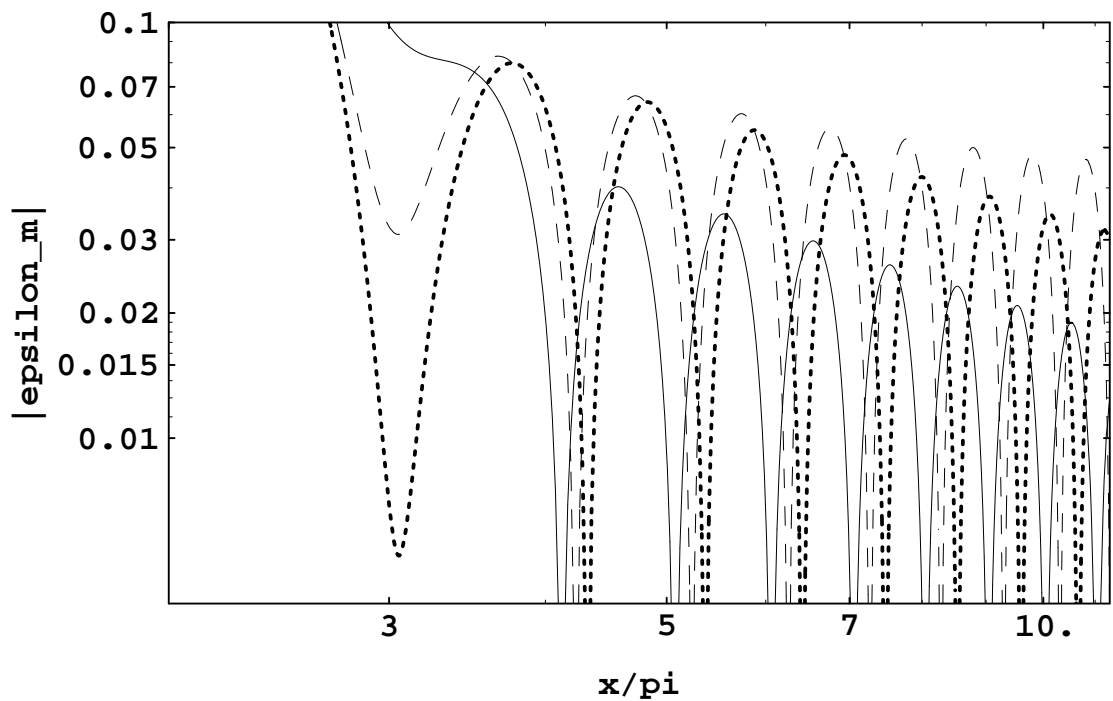} }}
\caption{In addition to the lines in Fig.\ 3, the dotted line
shows the  solution with the resummed kernel
($\lambda = 1$).}
\end{figure}

\end{document}